\documentclass[journal]{IEEEtran}
\usepackage{ifpdf}

\usepackage{cite}

\ifCLASSINFOpdf
  \usepackage[pdftex]{graphicx}
   \DeclareGraphicsExtensions{.pdf,.jpeg,.png}
\else
   \usepackage[dvips]{graphicx}
   \DeclareGraphicsExtensions{.eps}
\fi
\usepackage{xcolor}
\usepackage{blindtext}
\usepackage [autostyle, english = american]{csquotes}
\MakeOuterQuote{"}

\usepackage[cmex10]{amsmath}
\usepackage{algorithmic}

\usepackage{array}

\ifCLASSOPTIONcompsoc
  \usepackage[caption=false,font=normalsize,labelfont=sf,textfont=sf]{subfig}
\else
  \usepackage[caption=false,font=footnotesize]{subfig}
\fi
\usepackage{fixltx2e}

\usepackage{stfloats}
\fnbelowfloat
\ifCLASSOPTIONcaptionsoff
 \usepackage[nomarkers]{endfloat}
 \let\MYoriglatexcaption\caption
\renewcommand{\caption}[2][\relax]{\MYoriglatexcaption[#2]{#2}}
\fi
\usepackage{url}

\begin{document}
\title{A collaborative citizen science platform for real-time volunteer computing and games}
%
%
%

\author{Poonam~Yadav, 
        ~Ioannis~Charalampidis, 
        ~Jeremy~Cohen, 
       ~John~Darlington,
       ~and~Francois~Grey
\thanks{P. Yadav is with the Computer Laboratory, University of Cambridge, J. Cohen and J. Darlington are with Department of Computing, Imperial College London, UK. Contact e-mail: (poonam.yadav07@alumni.imperial.ac.uk, jeremy.cohen@imperial.ac.uk, j.darlington@imperial.ac.uk ).}
\thanks{I. Charalampidis is with CERN, CH-1211, Geneva 23, Switzerland. Contact e-mail: (ioannis.charalampidis@cern.ch) and F. Grey is with Citizen Cyberlab, University of Geneva, Switzerland. Contact e-mail: (francois.grey@unige.ch).}}

\markboth{IEEE TRANSACTIONS ON COMPUTATIONAL SOCIAL SYSTEMS,~Vol.~X, No.~X, November~2017}%
{Yadav \MakeLowercase{\textit{et al.}}: A collaborative citizen science platform for real-time volunteer computing and games}
%



\maketitle

\begin{abstract}
Volunteer computing (VC) or distributed computing projects are common in the citizen cyber-science (CCS) community and present extensive opportunities for scientists to make use of computing power donated by volunteers to undertake large-scale scientific computing tasks. Volunteer computing is generally a non-interactive process for those contributing computing resources to a project whereas volunteer thinking (VT) or  distributed thinking,  allows volunteers to participate interactively in citizen cyber-science projects to solve human computation tasks. In this paper we describe the integration of three tools, the Virtual Atom Smasher (VAS) game developed by CERN, LiveQ, a job distribution middleware, and CitizenGrid, an online platform for hosting and providing computation to CCS projects. This integration demonstrates the combining of volunteer computing and volunteer thinking to help address the scientific and educational goals of games like VAS. The paper introduces the three tools and provides details of the integration process along with further potential usage scenarios for the resulting platform. 
\end{abstract}

\begin{IEEEkeywords}
Citizen cyber-science, Online Games, Middleware, Cloud Computing, Volunteer Thinking, Volunteer Computing, Human Computation, Non-profit Sector, Crowdsourcing, Real-time Distributed Computing, Parallel Computing, Community Grid
\end{IEEEkeywords}

%

\section{Introduction}
%
%
%
%
\IEEEPARstart{I}{n}  the last few years, citizen cyber-science (CCS) has evolved as a new way of inspiring and supporting learning and participation in science. It provides a means for citizens, who may not have a scientific background, to interact with and contribute to scientific projects or studies. In many cases, such interactions are beneficial to both the scientist running the project and to the participating individuals who can gain new skills and knowledge in the process of supporting the project. Existing CCS projects are mainly categorised as volunteer computing~(VC) or volunteer thinking~(VT)  projects, examples of this are provided in~\cite{Crowdsourcing2013, Yadav2014Deployment, Yadav2016}. In a VT project, volunteers use their cognitive skill and knowledge to solve a part of a scientific problem; this type of project requires volunteers' active participation. In VC projects, volunteers contribute their computing resources to provide processing power to support one or more computationally intensive tasks within a project.  

Projects generally have specific use cases that are considered to be ideally suited to either volunteer computing or volunteer thinking. A project may have significant computational requirements that can best be served by farming out pieces of computation to the computers of volunteers who have opted to take part in the project. This allows passive volunteering from the user perspective with the user's computer becoming available to undertake computation for the project when the user is not actively using their machine. Alternatively, a project may be structured to take advantage of the ability of humans to undertake tasks that are computationally difficult but very straightforward for a human to process. Examples might include identifying particular properties of some data by looking at a graph or spotting visual anomalies in an image. Such tasks can be quickly and accurately undertaken by humans while reliably undertaking such tasks using code can be challenging and computationally intensive. These tasks are ideally suited to volunteer thinking where volunteers can actively participate in a project and assist the project owner in achieving their aims. Nonetheless, there are increasing numbers of use cases, particularly in the realm of scientific education, where volunteer thinking tasks can be made more realistic and more educationally valuable if they make use of realistic or pseudo-realistic data. For example, consider an online science teaching tool where real-world computation of scientific data that may be impractical to undertake on a single user's machine can be integrated into the project. The data generation is itself an ideal volunteer computing task and can be farmed out to the computers of one or more participants in the project to generate the required data to support a user interacting with the tool.

We therefore believe that integrating VT and VC into a single project can help to make interesting and engaging use cases for CCS projects. There are two main deployment scenarios for combining VC and VT within CCS projects. These have been presented in various CCS articles, an overview of which can be found in~\cite{ Yadav2014Deployment, Yadav2016a, Yadav2016}. The first scenario is where VT tasks are independent of VC tasks or where the two different types of task don't require real-time interaction with each other. The second scenario consists of a real-time interaction between the VT and VC tasks. In the first scenario where VC and VT tasks do not need to interact with each other in real-time or are completely independent, implementation is generally fairly straightforward and the two aspects of the project can be implemented separately, even though their implementations may ultimately be within the same codebase. The second scenario is more complex to implement since a common framework must be provided that can link the task lifecycles of both volunteer thinking and volunteer computing tasks. By this we mean that when a volunteer thinking task requires some data that must be generated to order, this must trigger a related volunteer computing task which must be scheduled such that results are provided to best support the interactivity requirements of the VT task the end user is undertaking. To the best of our knowledge there is currently no such collaborative platform described in existing literature and we refer readers to~\cite{Yadav2014Deployment} and~\cite{VAS2015} which provide an overview of this literature.

In this paper we detail the integration of three platforms, CitizenGrid~\cite{Yadav2015CitizenGrid,Yadav2016a}, LiveQ~\cite{VAS2014, LiveQ}  and the Virtual Atom Smasher~(VAS)~\cite{VAS2014, VAS} game, to show an example of a CCS project where VT and VC are integrated. In addition to this specific use case we look at how the integration of two of these platforms, CitizenGrid and LiveQ, can provide a more general platform for developing CCS projects and applications that integrate volunteer computing and volunteer thinking. In Section~\ref{Motivations} we discuss the motivations behind the integration of volunteer computing and volunteer thinking. In Sections~\ref{Platforms} and~\ref{Integration}, we first present a brief overview of the three target platforms followed by details of the design for the resulting integrated system. Sections~\ref{Implementation} and~\ref{UseCases} present implementation details and a set of usage scenarios for the integrated system. Section~\ref{Related} presents related work and we describe our conclusions and details of future work in Section~\ref{Conclusion}.

\section{Motivations}\label{Motivations}
The design, implementation and deployment of a citizen science project is driven by a goal which is first defined by considering the project creator's motivation and intentions.  The volunteers who participate in citizen science projects have different motivations and goals for their participation~\cite{Bowser2013}. However, the factors that primarily determine their decision between participating in a VC or VT project depend on their scientific and educational interests and the time and computing resources that they have available. Integrating VT and VC tasks into one project and allowing volunteers to participate in either or both types of task could make the project more popular and attract more participation. At the same time, integrating both volunteer thinking and computing into a project can significantly increase the complexity of building and running the project. Therefore, in this section we present our three main motivational goals for the integration of VT and VC in a single CCS project. \\

\subsection{Volunteer Engagement}
Engaging and retaining volunteers in a project is a challenging task. In recent years, a number of research studies have been conducted to understand what motivates volunteers and what project factors influence their continued participation~\cite{Bowser2013, Jennett2014, Charlene2013a,Darch2010}. It is understood from various studies that gamification of VT tasks\footnote{The use of gaming approaches to represent scientific human-computation tasks.} keeps people more entertained and engaged in the project~\cite{Bowser2013, Shahri2014, Zhao2014}. Therefore, incorporation of serious VT games with VC projects, that generally have limited interaction, will offer another approach to enhance user participation in VC projects and make them more rewarding to take part in.\\

\subsection{Scientific Education}
Gamification and VT task interfaces provide great learning and creativity opportunities for volunteers.  CCS game players and volunteers learn and enhance their project specific scientific knowledge through hands-on experience with real tools.
For example, in the Virtual Atom Smasher game, the players learn about particle physics by visualizing simulated results that are generated using equation parameters that the volunteers can alter. The simulations correspond to the real-world experiments being undertaken by physicists at CERN. The game players also learn about various new computing technologies, for example, how to work with the VirtualBox~\cite{Virtualbox} virtualization software. Without the integration of volunteer computing, the computation that would be required to undertake the simulations may be too much for an individual user's computer system. This would either mean that the generation of results would take too long and spoil the game-playing experience or that result generation using real-world parameters would not be possible at all.\\

\subsection{Participation Diversity}
Previously, volunteer computing projects have attracted only volunteers who wish to contribute spare computing resources to a project but are not so interested to interact with the project directly. The volunteer computing platforms~\cite{Yadav2015CitizenGrid, WCG2015} offer the ability for volunteers to gain straightforward access to a range of citizen science projects registered with them. This, combined with projects that integrate both volunteer thinking and volunteer computing processes, opens up a range of opportunities to attract new volunteers who have time to play games and are also interested to participate in and gain understanding and knowledge of scientific processes and challenges. We categorize volunteer participation in two levels: single-mode participation (either computing or thinking only), or combined participation.  With single-mode participation, while a project may integrate both volunteer thinking and volunteer computing, it is not necessary for a volunteer to participate using both approaches to take part in the project. However, in combined participation a volunteer has to use both approaches to take part in the project. In the VAS-CitizenGrid scenario, while the project integrates aspects of volunteer computing and volunteer thinking and both are required for the game to operate successfully, individual game players need not contribute their computing resources for the volunteer computing tasks but they can still participate in the learning aspects of the project. Game players can take part in the project as part of a team where some individuals in the team may participate only as volunteer computing providers, others may participate from the volunteer thinking perspective. It is also possible to be both a volunteer thinking and volunteer computing participant -- in this case the user is considered to be providing combined participation. If game players are not contributing their own processing resources when playing the VAS game, their computing tasks can be processed by other team members' computing resources. A survey has shown that less than 10\% of participants in volunteer computing projects are women, whereas, in online games, the participation of women and men is nearly equal~\cite{Raoking2014, Estrada2013}.  The survey results further support our strong motivation to integrate scientific gaming with VC projects to attract new participants to projects and ensure the involvement of a diverse range of volunteers.

\section{Background}\label{Platforms}
In this section we provide a brief overview of each of the platforms used in this work.

\subsection{Virtual Atom Smasher~(VAS)}

VAS~\cite{VAS2014, VAS, VAS2015a} is a web-based physics game, developed at CERN,  that aims to educate game players about particle physics through an interactive hands-on web-based environment.  The game is based on the concept of a ``Virtual Collider'' that represents a real particle collider. The game front-end is developed as a web application. The interface contains interactive videos, rich pop-up explanations and animations, which help players to understand the particle physics fundamentals, for example, high energy particle collision. We use a set of screenshots from the web-based interface of the VAS application~\cite{VAS2015a} in this section to give the reader an idea of the highly graphical nature of the game and how it presents scientific data and concepts to game players. 

\begin{figure}[h]
 \centering
\fcolorbox{black}{white}{ \includegraphics[width=65mm]{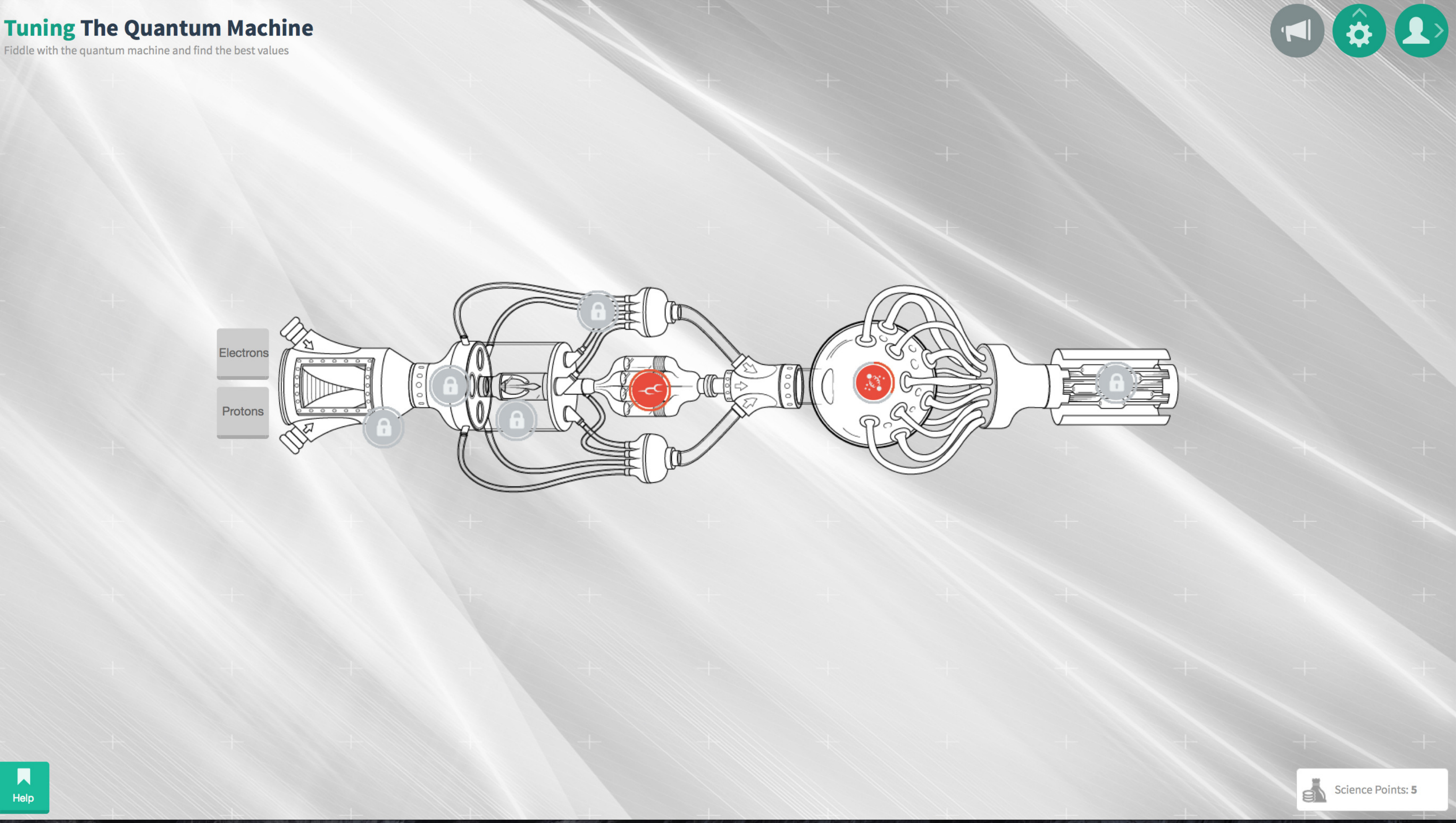} }
\caption{Virtual Atom Smasher Game "Quantum Machine" Interface.}
\label{fig:VASgui} 
\end{figure}

\begin{figure}[h]
 \centering
\fcolorbox{black}{white}{ \includegraphics[width=65mm]{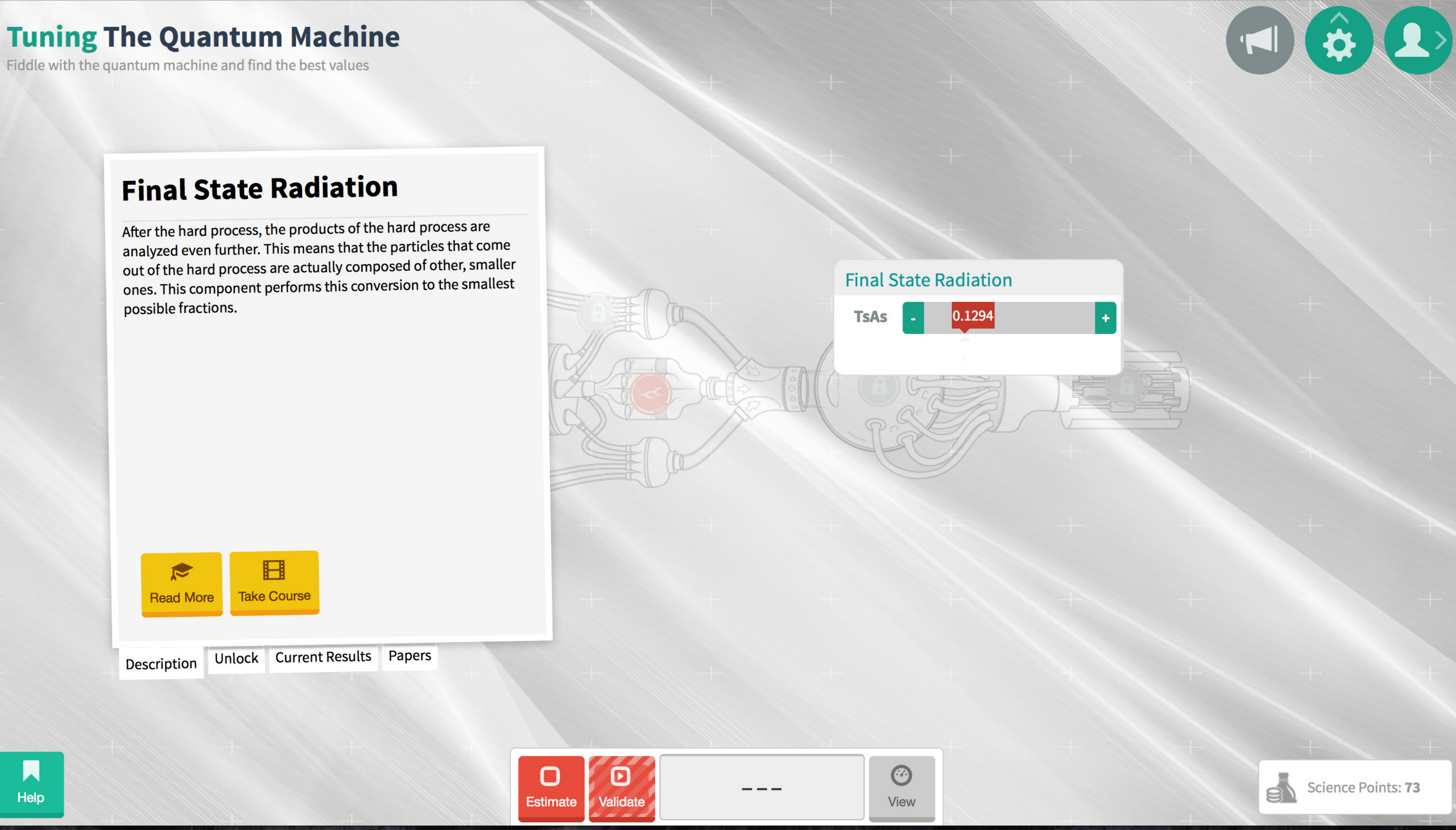} }
\caption{Virtual Atom Smasher game interface showing the parameter setting and validation.}
\label{fig:VASparameters} 
\end{figure}

In Figure~\ref{fig:VASgui}, the  game user interface shows a ``quantum machine'' with a number of locks, in principle it's an attempt to visualize the sequence of events that occur inside the event generation software which simulates a particle collision inside the Virtual Collider. The game player's goal is to unlock all the locks on the machine.  In order to unlock parts of the machine they need to spend science points that they have earned during the game. They earn science points by choosing and validating suitable tuning parameters (See Figure~\ref{fig:VASparameters}). 

The VAS game back-end is developed to tune Monto Carlo simulations of high energy particle physics.  The common problem in such cases is to find the correct values of a dozen parameters such that the simulation results match the observed results from current and past particle colliders. Even though, for this classical minimization problem, there are already automated solutions (e.g.~\cite{Professor2015}) we considered that the educational value of an interactive interface was much greater than using automated approaches\footnote{This is one of the reasons why CERN has begun to grant access to the real scientific software and data that physicists use to anyone interested in the science of particle physics, via the portal http://opendata.cern.ch/.}. The game front-end is developed as a web application using the latest HTML5 technologies for enhancing the educational experience and providing players with a step-by-step approach towards understanding the world of particle physics.

\begin{figure}[h]
 \centering
\fcolorbox{black}{white}{ \includegraphics[width=65mm]{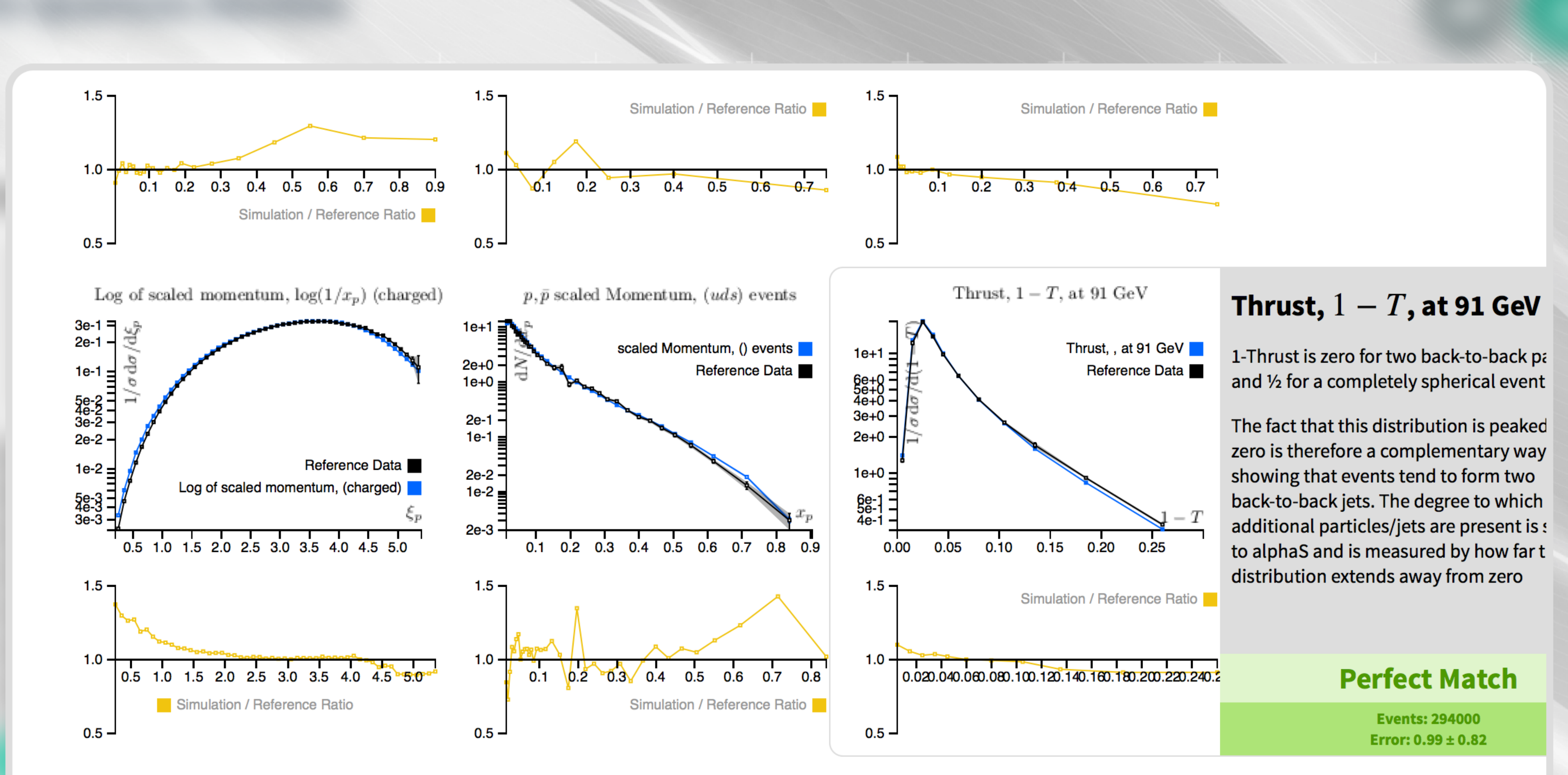} }
\caption{Virtual Atom Smasher game interface showing simulation histograms.}
\label{fig:VAShistogram} 
\end{figure}

When players select and validate a parameter on the game interface, they request the execution of a Monte Carlo simulation to a set of distributed computer nodes.  A player's task is to fine-tune the parameters of the CERN Virtual Collider simulator in such a way that the simulation presents the same results as those observed in the real experiments.  The number of  "science  points" or "credits" a player gets depends on two factors. First, the values of the parameters they choose for the simulation. Generally, the first value chosen within a provided range is a random guess. However, in subsequent steps they can adopt a more guided approach, making decisions based on results provided by their previous chosen values. Figure \ref{fig:VAShistogram} shows a group of histograms that provide the results of previous simulations. The second factor for gaining credits is the speed at which simulations are processed. By processing their simulations more quickly, a user can try and validate more guesses at possible input values. If they provide a larger number of computing resources to run their simulations, they can undertake more simulations in a given period of time enabling them to gain more credits.

The Virtual Atom Smasher game provides many opportunities for those interested in learning about particle physics to gain some understanding of the field through a fun and competitive game-based environment. The integration of volunteer computing and volunteer thinking in this environment offers an incentive for more volunteers to take part in the game and we believe that the approach used offers opportunities to provide a general educational environment for those interested in taking part in a variety of CCS games.

\subsection{LiveQ}

LiveQ is a job distribution and monitoring middleware with real-time interaction capabilities for volunteer computing projects. Generally, volunteer computing projects are based on a distributed Client-Server model~\cite{Yadav2014Deployment} for VC task distribution and management.  Instead of designing the distribution framework from scratch, many VC projects use generic middleware platforms that are specifically designed and implemented to support Volunteer Computing.  For example, the SETI@Home project~\cite{Seti}, a popular volunteer computing project, uses the BOINC~\cite{Boinc,Yi2011} middleware for  its task distribution and management.

LiveQ offers real-time feedback and control of tasks and simplifies the process of managing large numbers of tasks and the integration of their results.  LiveQ maintains a record of all the volunteer computing resources connected to it any time, which makes it a suitable choice for a game-based integrated project. The VAS game uses LiveQ for its task management. 

\begin{figure}[h]
 \centering
\fcolorbox{black}{white}{ \includegraphics[width=65mm]{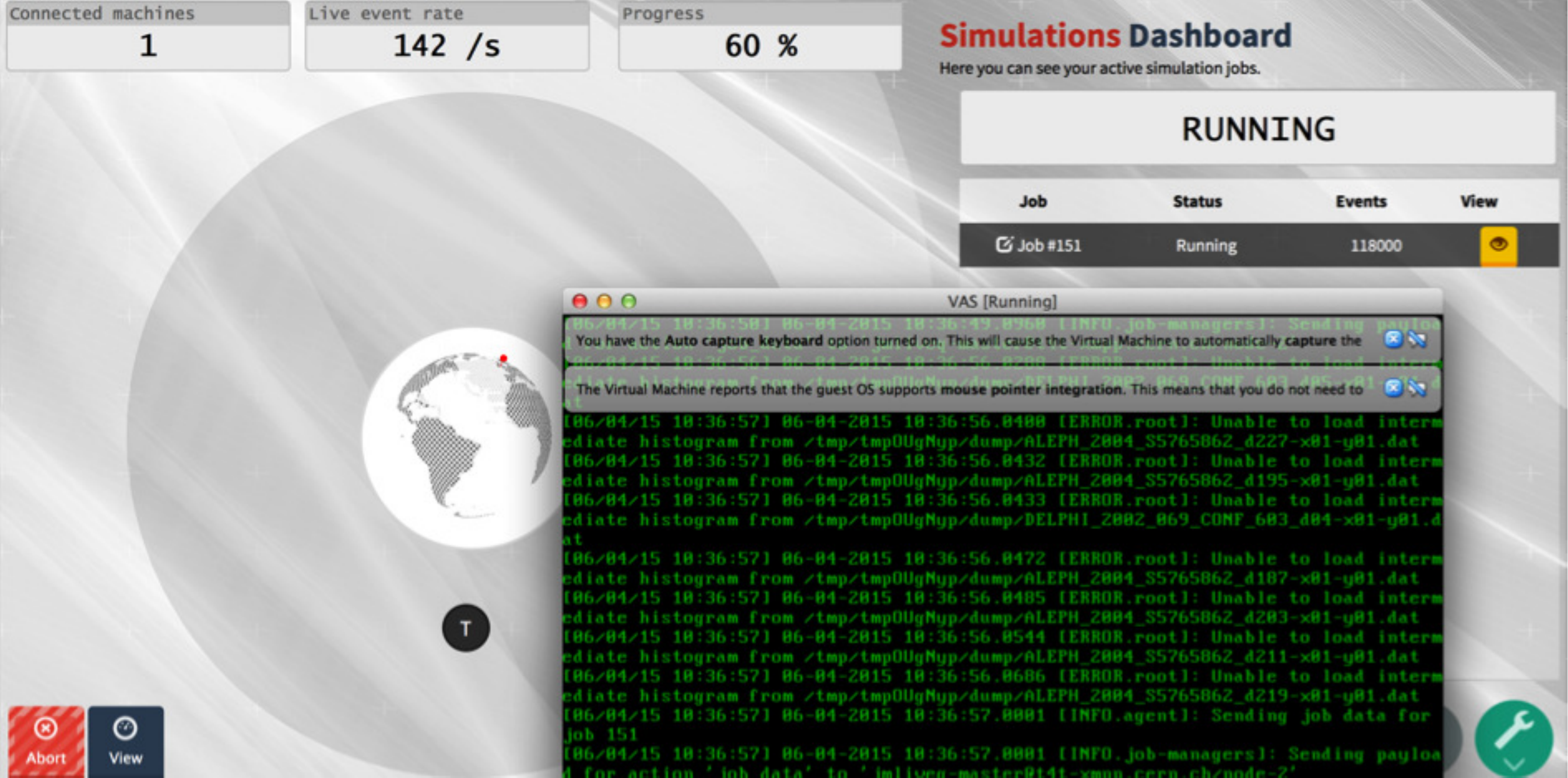} }
\caption{Virtual Atom Smasher game interface showing real-time volunteer computing task progress status.}
\label{fig:VASstatus} 
\end{figure}

Figure~\ref{fig:VASstatus} shows the VAS game interface displaying realtime task progress status. It also shows the number and details of volunteer computing client machines that are undertaking processing of tasks on behalf of the game player currently logged in to the interface.

\subsection{CitizenGrid}

CitizenGrid is a web-based platform that provides the hosting, deployment and management of CCS applications~\cite{Citizengridgit}. Scientists or CCS project managers can deploy server images for their applications onto different cloud platforms or onto local server resources and also register their application client images with CitizenGrid so that they are available to end-users (volunteers) who want to participate in CCS projects.
 
For volunteers, CitizenGrid provides a CCS application portal where they can discover and participate in CCS applications by launching application clients on their local machine, or where applicable, on a remote cloud platform, for example, OpenStack~\cite{OpenStack} or the Amazon EC2 cloud infrastructure~\cite{AWS}. 

\begin{figure}[h]
 \centering
\fcolorbox{black}{white}{ \includegraphics[width=65mm]{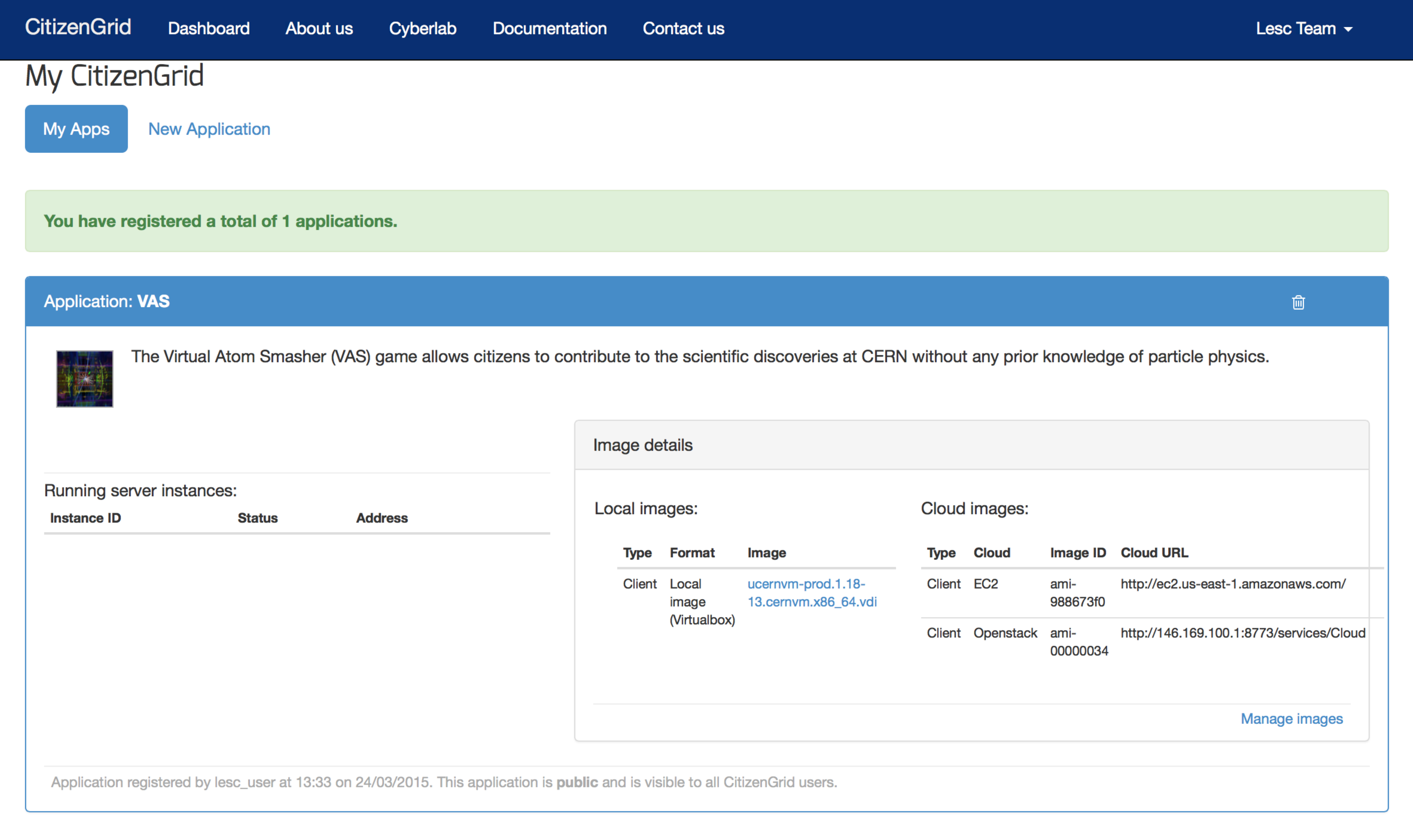} }
\caption{CitizenGrid interface showing details of the VAS game as a registered volunteer computing project.}
\label{fig:CGprovider} 
\end{figure} 

While CitizenGrid is intended to be a generic platform, complex citizen science applications such as VAS can be better supported and provide an enhanced user experience by adding custom application-specific extensions to CitizenGrid. The integration of VAS, LiveQ and CitizenGrid provides enhanced accessibility and usability of the VAS game, building on the capabilities of LiveQ and CitizenGrid to support deployment, management and the integration of volunteer thinking and computing.

In the following sections we provide an overview of the integration process used to link VAS, LiveQ and CitizenGrid and present implementation details of the resulting platform, its potential use cases and related work. 

\section{Platform Integration Overview}
\label{Integration}

We refer to the integration of the Virtual Atom Smasher (VAS) citizen science application~\cite{VAS2014}, the LiveQ job distribution framework~\cite{VAS2014, LiveQ} and the CitizenGrid~\cite{Yadav2015CitizenGrid, Citizengridgit} platform as VAS-CitizenGrid. In this section, we provide an overview of the structure of the integrated environment then detail the participation and collaboration that needs to be undertaken by individuals working within the VAS-CitizenGrid environment.  

The VAS game provides an interface for individuals to learn about particle physics but it requires underlying computation to support this process. While volunteers who want to participate only from a volunteer thinking perspective can visit the game website and begin working through the game and learning about the physics aspects of the ``Virtual Collider'', their interactions will trigger the running of computations and these need to be carried out somewhere. These computations could be carried out on central resources run by the game's operators but these computations can be large and this is an unsustainable approach as user traffic grows and more people want to take part. Other options might, for example, include using Infrastructure-as-a-Service (IaaS) cloud resources to support the scaling of resource capacity. However, using such publicly available platforms requires that computation is paid for on a per-use basis. For an educational tool that is freely accessible, this is also not a practical solution. Volunteer Computing is an ideal solution for such a task.

VAS uses the LiveQ job distribution framework to enable and manage the distribution of jobs to the computing resources of volunteers. LiveQ deals with the process of registering users' resources and keeping track of them so that tasks can be farmed out to these resources when user interactions within the VAS game generate them. This explains the reasoning for the integration of VAS and LiveQ, however, there is still a challenge. Volunteers who want to provide their computing resources need a way for these resources to be enabled to run the VAS simulation software which can undertake the jobs that are farmed out by LiveQ. In the case of some volunteer computing projects, the computation done on volunteers' computers is relatively simple. This doesn't necessarily mean that the software is not computationally intensive but it does mean that the code is sufficiently simple that a single small executable, perhaps with a small number of dependencies, can be downloaded to a volunteer's resource and run. This generally makes distributing the code very straightforward. In the case of the VAS simulation software, the code is rather more complex requiring a specific operating system and stack of dependencies and complex configuration. The CernVM virtual machine appliance~\cite{CERN} was developed specifically to address this challenge and is described in more detail in Section~\ref{section:integration-cg-liveq}. For volunteers to make use of the CernVM virtual machine, it must be deployed and run on their resources and configured to tell the LiveQ infrastructure which group a user is a member of. This ensures that only computations belonging to that group are distributed to the user's resource(s). This is where CitizenGrid provides an ideal environment to link users with the integrated VAS and LiveQ framework. Users register with CitizenGrid which provides group management allowing users to form themselves into groups that can then be used within the VAS game. CitizenGrid also provides the ability to host the CernVM machine image and deploy and run this on a user's local machine in a straightforward manner. This can be done using either CitizenGrid's legacy setup for starting VirtualBox virtual machines based on Java Web Start technology or using the CernVM Web API~\cite{VAS2015} that provides an enhanced method of starting and managing virtual machines running in VirtualBox. CernVM Web API has been integrated within CitizenGrid which allows enhanced management of CernVM virtual machines running in VirtualBox from within the CitizenGrid web application.

The linking of these three platforms therefore provides a full end-to-end system allowing end users to find the VAS application within the application directory in CitizenGrid, join a group or team, and then start one or more virtual machines capable of running the required CERN software stack to enable these machines to undertake simulations of data used in the VAS game. Once a virtual machine is running, it registers with the LiveQ infrastructure which then adds it to the list of machines available to undertake computations for the target group. The volunteer may then play the VAS game themselves or provide their compute resource(s) for others in the same group who are interacting with the game. When computations are generated by game players in this group, they are farmed out to available resources and run, providing feedback to game players within the VAS user interface.

\begin{figure}[t]
 \centering
\fcolorbox{black}{white}{ \includegraphics[width=65mm]{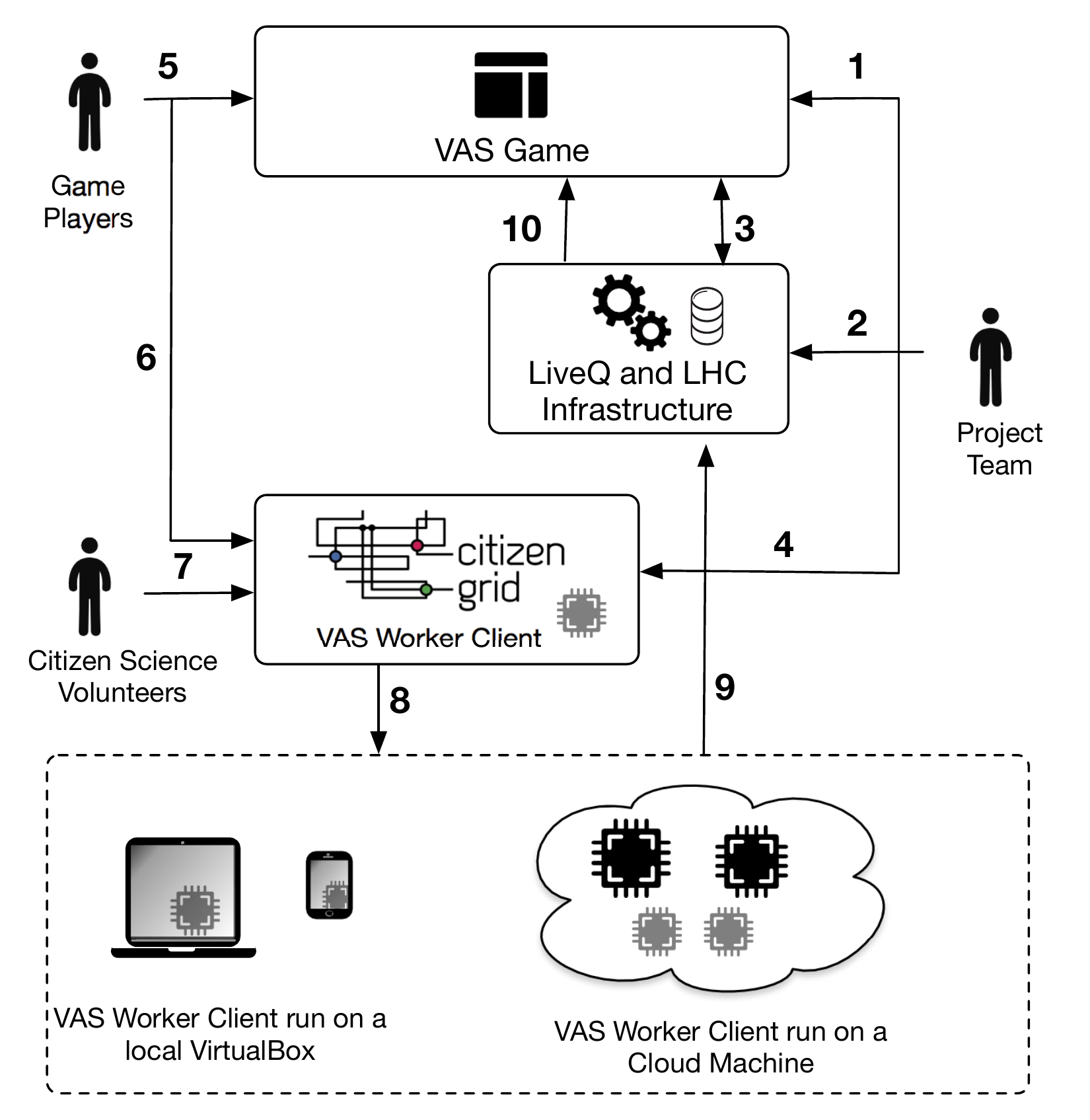} }
\caption{VAS-CitizenGrid collaborative platform. Numbers in the figure show participation and interaction steps.}
\label{fig:OverviewFigure} 
\end{figure}

Figure~\ref{fig:OverviewFigure} shows the ordering of the interactions between VAS, LiveQ and CitizenGrid. We now describe the different participants within such an integrated environment and how they interact with each other. While these descriptions focus on the VAS-CitizenGrid exemplar, they would also apply in the context of other integrated volunteer thinking and volunteer computing applications. \\

\textbf{Project Participants}\\

A CCS project, such as VAS, that uses integration with CitizenGrid and LiveQ to provide its operational platform to make games available and deploy VC tasks,  involves a number of participants. We categorise these participants as project creator teams, game players and CCS volunteers:

\begin{itemize}
\item \textbf{Project Creator Team}
\begin{itemize}
\item \textbf{Scientists}:  Scientists define scientific game requirements, specifications, descriptions, and designs.  They provide  scientific data, procedures, and the methods that support volunteer computing tasks.\\
\item \textbf{Educators}:  Educators develop the educational material based on the scientific processes involved and refine game designs from the learning and creativity perspective.\\
\item \textbf{Developers}: The developers may include individuals from a range of different development roles -- server-side/back-end developers, mini-game developers, UI developers, etc. These developers implement and test the various software components of the game and the VC tasks that will be deployed to volunteers' computers.\\
\item \textbf{CitizenGrid Application Provider}:  The application provider is the individual in the project creator team who registers the project with the CitizenGrid environment by providing project details and uploading the project's volunteer computing client virtual images (worker nodes) that will be downloaded by volunteers' computers. 
\end{itemize}
\item \textbf{Thinking Volunteers - Game Players}: These are the individuals who play a scientific game in order to both learn about the scientific processes or domain represented in the game and to support the game operators in undertaking some scientific task.
\item \textbf{Computing Volunteers - Citizen Science Volunteers}:  Computing volunteers contribute to the CCS project by donating their computing resources. This is achieved by downloading and running VC client images on their local computer. Alternatively they may choose to support the project by providing their own cloud infrastructure resources or funding remote, public cloud resources, that run the project's volunteer computing images. \\\\
\end{itemize}

\begin{figure}[h]
 \centering
\fcolorbox{black}{white}{ \includegraphics[width=65mm]{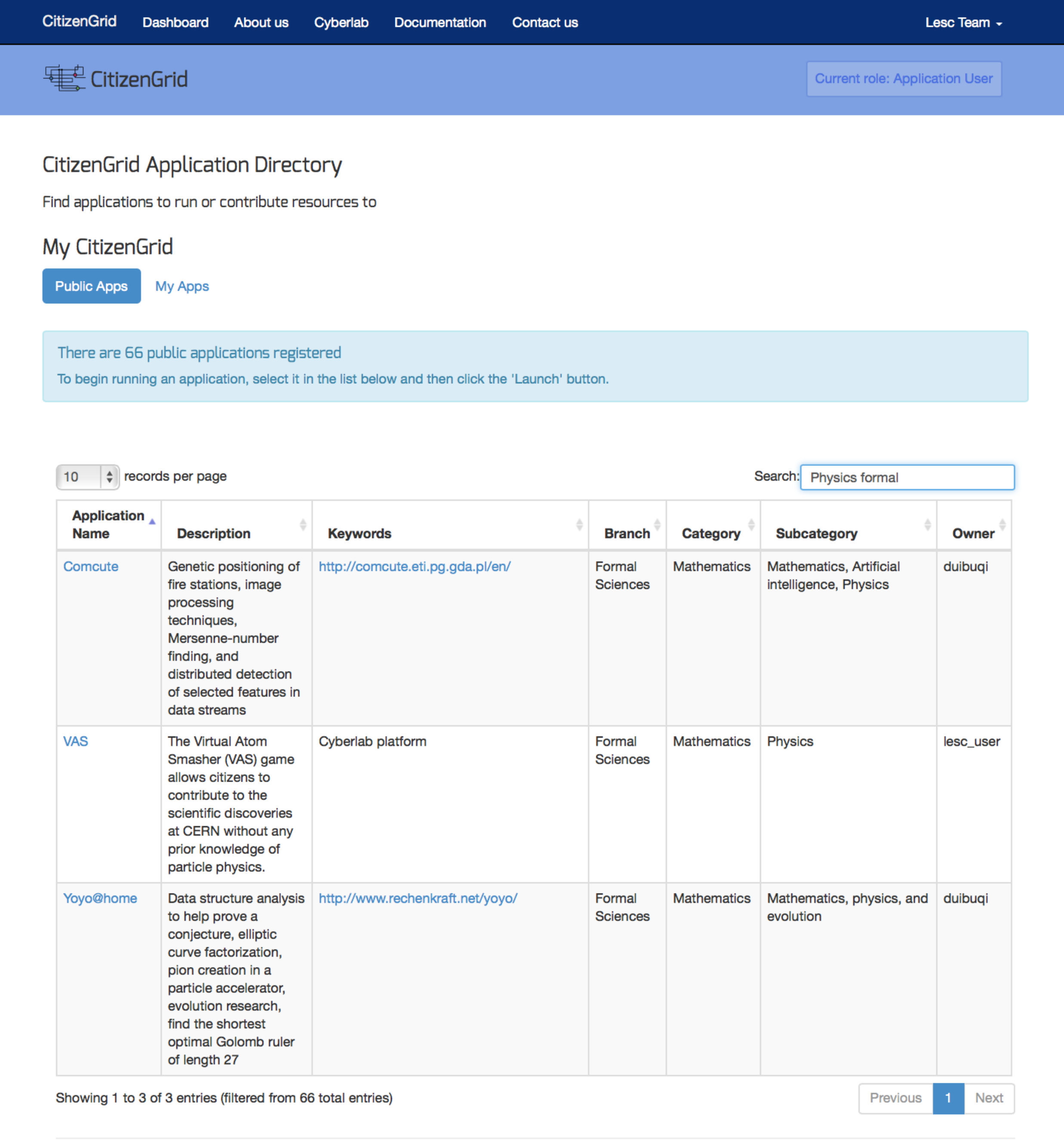} }
\caption{CitizenGrid application directory showing an example of available CCS projects that users can choose to participate in, including examples of third-party publicly available CCS projects from other groups.}
\label{fig:CGuser} 
\end{figure}

Note that an individual may fall into more than one of the above roles when participating in an integrated VT/VC project.\\
\noindent

\textbf{Participation and Interaction Steps in the VAS-CitizenGrid platform}\\\\
We now look at the steps shown in Figure~\ref{fig:OverviewFigure} detailing the collaboration between the project creator team, game players and citizen science volunteers. The steps in the diagram are numbered and the corresponding descriptions are provided below:

\begin{enumerate}
\item The project team creates and deploys scientific learning games.
\item The project team sets up a volunteer computing (VC) infrastructure using LiveQ for the real-time distribution and management of VC tasks.
\item The project team configures game and VC components to interact with each other.
\item The project team registers/hosts the VC client  images (worker nodes) on the CitizenGrid platform.
\item The game players register their team with the VAS game via the VAS web-based interface and are assigned a team identifier.
\item The game players create a team with their VAS team identifier on CitizenGrid.
\item The volunteers register with the CitizenGrid platform and join a team.
\item The volunteers participate in the volunteer computing project by downloading the VC worker node to their local computer and running it via the VirtualBox virtualization software, or by running a worker node on a cloud infrastructure.
\item A running worker node receives tasks directly from the VC infrastructure (Server).
\item As a game player interacts with the game interface, the VC server updates the interface with new game information and generates new VC tasks to be sent to volunteers' computers for processing.
\end{enumerate}

\section{Implementation} \label{Implementation}
The particular scenario of integrating volunteer thinking and volunteer computing is enabled in the system described in this paper by linking the CitizenGrid, LiveQ and VAS platforms. As in the previous section, we again refer to CitizenGrid-VAS as the collaborative platform that combines the functionalities of CitizenGrid, LiveQ and VAS. By replacing the VAS game with other citizen cyber-science games or applications which require real-time interaction between VT and VC tasks, the platform has the ability to offer support for a variety of CCS use cases.  In this section, we provide technical details of the integration of the platform components in the CitizenGrid-VAS environment.

\begin{figure}[h]
 \centering
 \fcolorbox{black}{white}{\includegraphics[width=65mm]{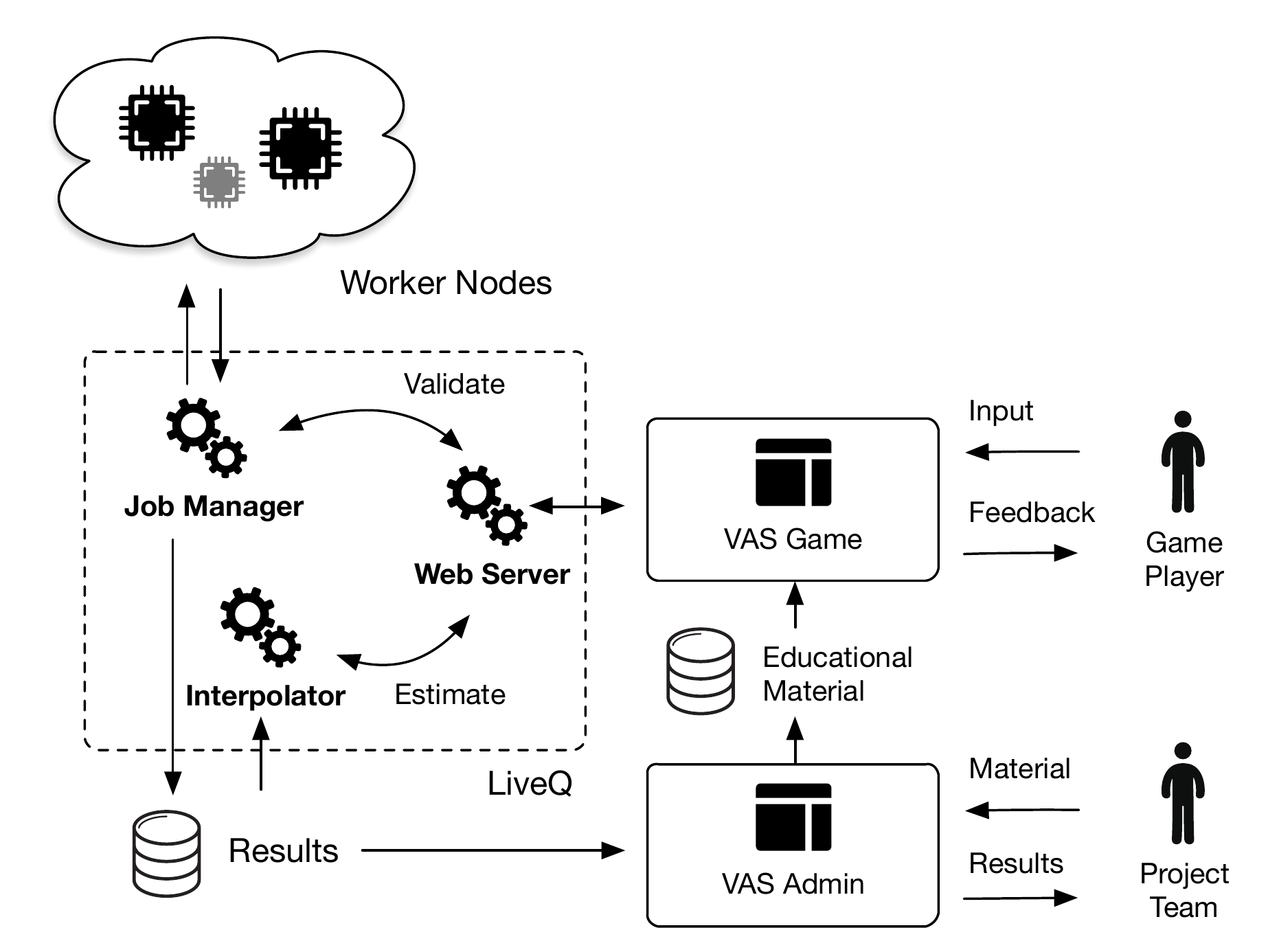} }
\caption{Virtual Atom Smasher game and LiveQ interactions.}
\label{fig:VAS} 
\end{figure}

\subsection{Integration between the VAS Game Interface and LiveQ }
The integration of the VAS game and the LiveQ framework is shown in Figure~\ref{fig:VAS}. When the game player chooses a parameter and clicks on the "validate the parameter" button, the backend functionality requests the execution of a Monte Carlo simulation. These simulations will be carried out by one or more distributed computer nodes. This request is routed through the LiveQ framework, which handles the control of the connected volunteer computing resources and overseeing the simulation process. The framework is designed in such a way as to minimize the overall run time and to keep the user informed throughout the simulation process, providing as much information as possible. This is an important requirement to keep the overall experience interactive since every simulation can take up to half an hour to complete. 

The LiveQ framework~\cite{LiveQ} uses a multivariate interpolation mechanism to estimate the outcome of the simulation and send a "guess" to the user within a few seconds of their request. In the meantime, it dispatches the job to as many volunteer computing worker nodes as possible. The LiveQ framework is fully aware of the entire network of worker nodes, and therefore it can pick the most appropriate ones, or force some of them to discard their current work and start a new task. The job-manager component communicates with the worker network using the Jabber/XMPP protocol that offers flexibility and scalability, even in environments with a slow network or firewall restrictions.

The simulation process used in VAS is highly parallelisable and it is very easy to merge results as they arrive from different worker nodes meaning that the process can easily be handled by multiple independent, distributed computing resources. This is particularly important since the use of real scientific data-sets means that significant computation may be required and this can be speed up through the use of larger numbers of compute nodes. Therefore, after a fixed number of events, each worker node sends back its intermediate results, gradually optimising the results presented in the GUI. When all the workers have completed their tasks, their results are merged into a final results record. The LiveQ framework also takes care of comparing these results against the experimental results and giving a ``Goodness of fit'' score~\cite{Smyth2003}. The LiveQ framework calculates the $X^2$ test score between the histograms produced by the simulation and the histograms obtained by the experiments. Therefore, it is possible for the system to automatically understand if the user has succeeded in finding a good solution, and to give the appropriate credit. 
\begin{figure}[h]
 \centering
\fcolorbox{black}{gray}{ \includegraphics[width=65mm]{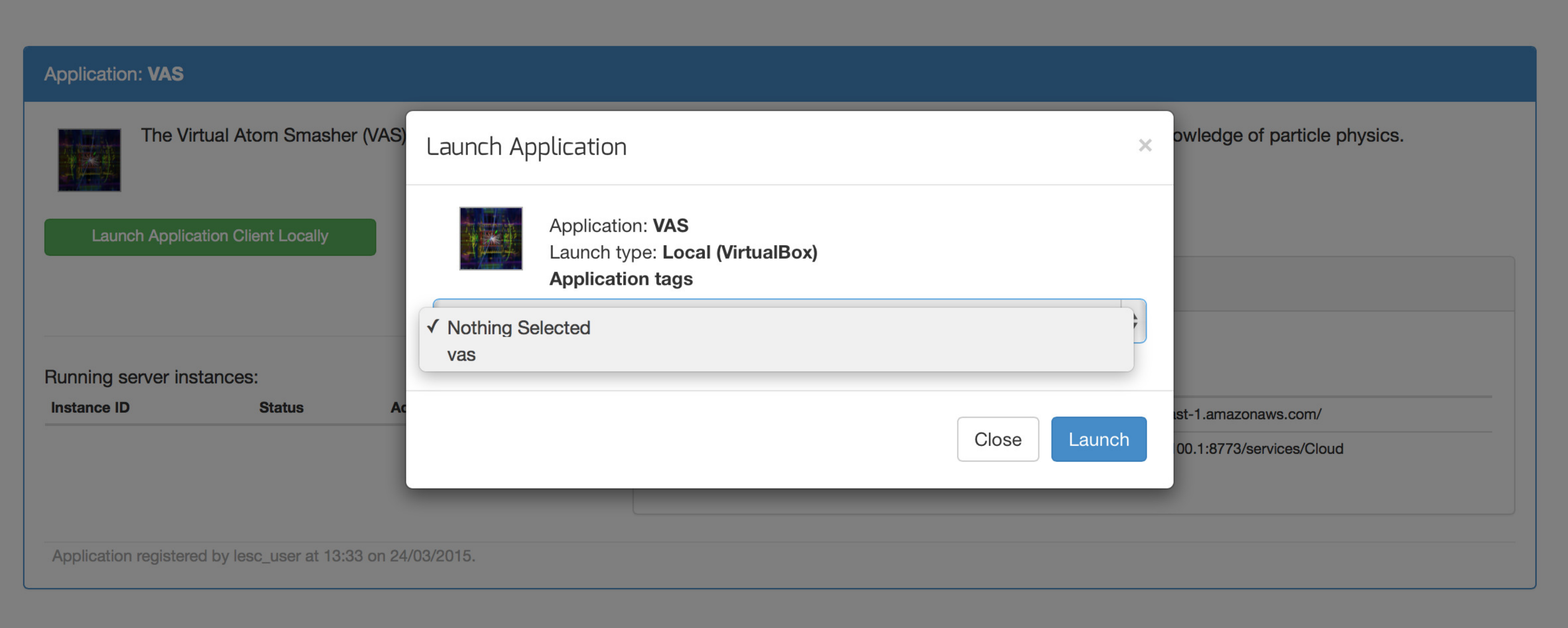}} 
\caption{CitizenGrid interface showing the VAS worker node launch using the team-alias.}
\label{fig:CGapplaunch} 
\end{figure}

\subsection{Integration between CitizenGrid and VAS}
CitizenGrid~\cite{Yadav2015CitizenGrid} has been implemented using the Django framework~\cite{Django}. CitizenGrid is intended to provide a unique one-stop environment for volunteers to search for their favourite volunteer computing and gaming projects and to set up and manage game teams through which they would like to contribute their resources. For example, a volunteer signs up to the CitizenGrid platform in order to begin taking part in a citizen cyber-science project. For team-based projects, the user needs to select their chosen project and become part of a team. There are two possible approaches to doing this: either they find and select their favourite project and then join a team that is already participating in that project or they first join a team and then select a project to participate in from those that the team is involved with. This creates a resource pool for the team. The advantage of contributing resources through CitizenGrid is that volunteers have an option of contributing their resources to many projects at the same time and can use a common environment to manage their project participation. In addition to their local resources, volunteers can also contribute resources from a public cloud infrastructure. 
\begin{figure}[h]
 \centering
 \fcolorbox{black}{white}{\includegraphics[width=65mm]{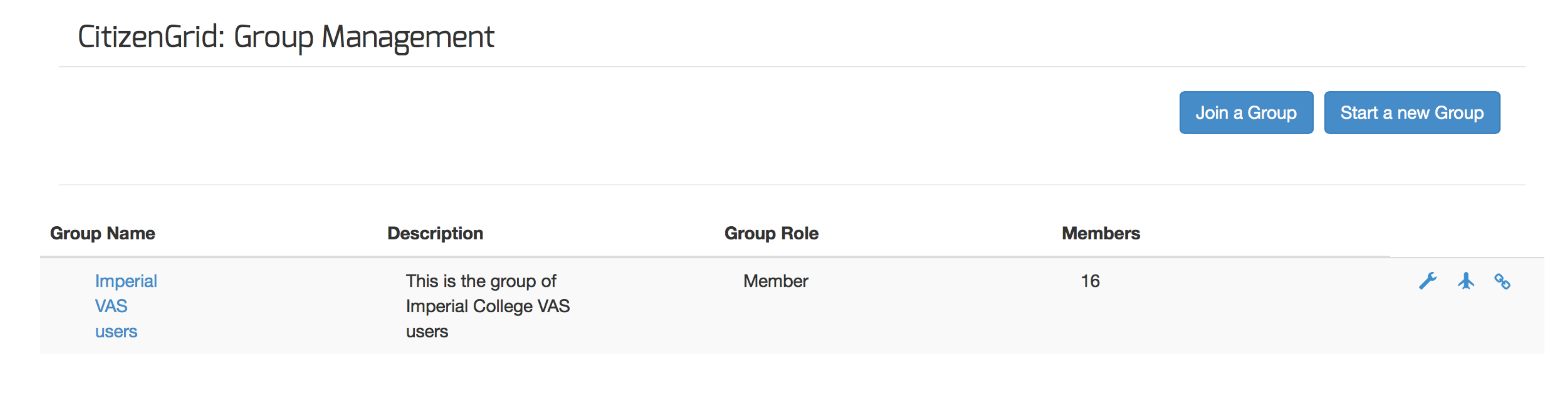} }
\caption{CitizenGrid "Group management" interface showing the list of existing groups, and options of starting a new group and joining an existing group owned by another volunteer.}
\label{fig:CGgroup} 
\end{figure}

\begin{figure}[h]
 \centering
 \fcolorbox{black}{gray}{\includegraphics[width=65mm]{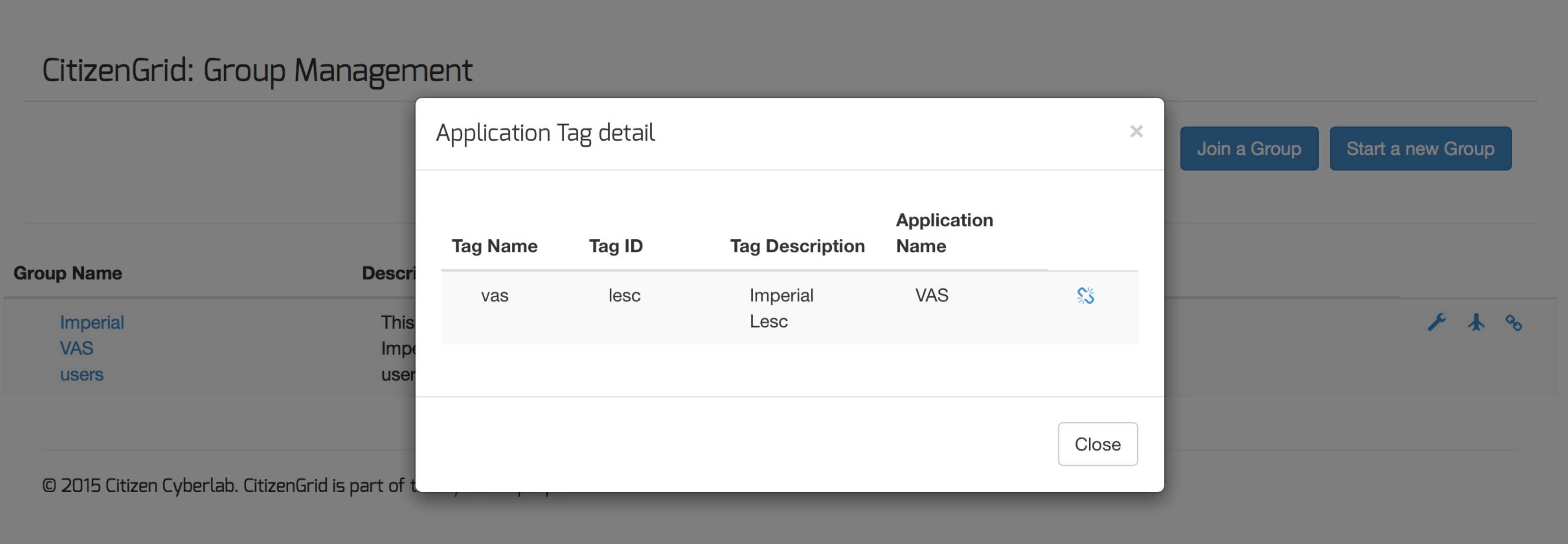} }
\caption{CitizenGrid "Group management" interface showing the VAS application "attachment" to the group using the team-id (labeled as Tag ID).}
\label{fig:CGapptag} 
\end{figure}

\begin{figure}[t]
 \centering
\fcolorbox{black}{white}{ \includegraphics[width=65mm]{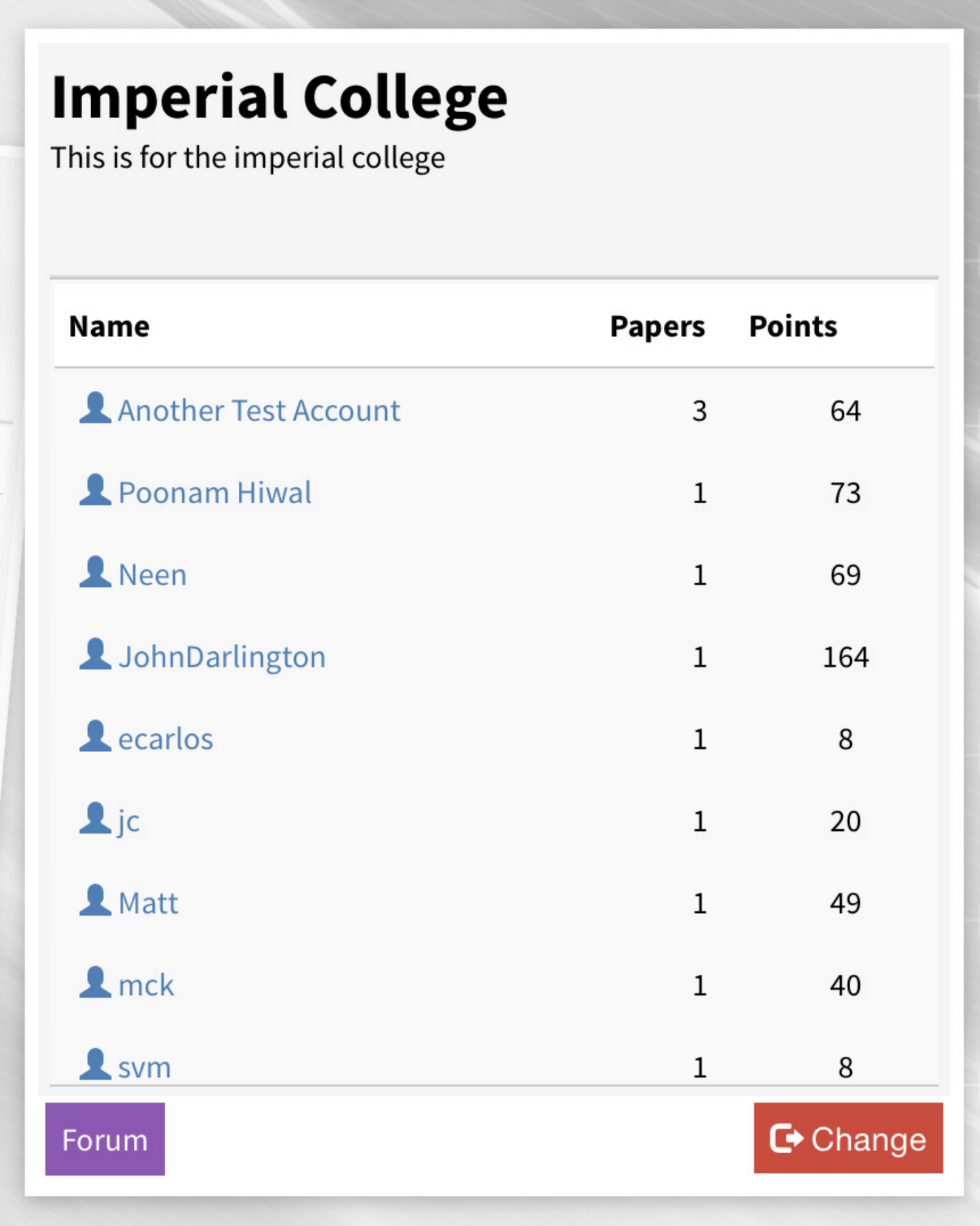}} 
\caption{The VAS interface showing game players within a team and the credits earned by the computation carried out by the CitizenGrid team's computing resources.}
\label{fig:VAS_team} 
\end{figure}

In the case of the Virtual Atom Smasher game, each player belongs to a team and the team uses computing resources donated to it for running simulations. The game players use CitizenGrid as a middleware for requesting the computing resources for their team from the volunteers that are registered with CitizenGrid. VAS and CitizenGrid link game players' worker nodes to volunteers' resources using a \textit{team-id}. This is a unique identifier for a team that is shared between game players and volunteers. CitizenGrid provides a feature for game players to create a group and then attach the VAS application to this group using their unique VAS \textit{team-id} and  \textit{team-alias} (see Figure \ref{fig:CGgroup}  and  \ref{fig:CGapptag}).  Once this has been done, other volunteers joining the same group can see the group's details including the attached VAS application. When volunteers launch the VAS application, they  see a drop-down menu for selecting  a \textit{team-alias} as shown in Figure~\ref{fig:CGapplaunch}. Selecting a team alias from the list results in the contextualisation of the VAS worker node (see Section~\ref{section:integration-cg-liveq}) so that it registers itself as belonging to the specified team and is therefore made available to undertake computations for that team. Figure~\ref{fig:VAS_team} shows the VAS interface displaying details of a team including the list of team members and the credits they have earned.

The integration of VAS and CitizenGrid was enabled through the addition of group creation and management features that are compatible with the approach used by VAS. We consider that these features are a valuable addition to CitizenGrid since they're likely to be applicable to other volunteer computing and volunteer thinking applications that providers may want to register with CitizenGrid.

\subsection{Integration between CitizenGrid and LiveQ}\label{section:integration-cg-liveq}
Because of the complex stack of dependencies that the experimental software run by VAS requires, it is not possible to cross-compile the software for every operating system. Therefore, the VAS project team decided to use virtualisation for  worker nodes, using the micro-CernVM~\cite{CERN} as the  base virtual appliance. This small virtual machine image provides the basis for a tailored OS distribution for volunteer computing projects because it minimises the overall data footprint by transferring the absolute minimum required data over the network. Micro-CernVM is derived from the full CernVM~\cite{CERN} virtual appliance, which is a well-established distribution for experimental software. One of the big advantages for using CernVM is its file system (CVMFS)~\cite{ref:cvmfs} that offers a high-performance remote file store providing a reliable way of delivering software and data to distributed computer nodes. Nodes mount the CVMFS filesystem and file data is only transferred across the network to the compute nodes when files are requested. This ensures that the absolute minimum required data is transferred across the network to the compute node. Finally, CernVM does not require modifications to the base image to run a given application. Instead it uses a ``contextualization'' mechanism to define its boot behaviour. This contextualisation process determines the functionality that a CernVM instance will provide when it starts up. The required software, already deployed in the  CernVM File System, can be pulled to the instance at startup time. The worker nodes make use of the LiveQ agent scripts that are already available in CVMFS, and are directed to provide resources for a particular group through the contextualization process.  CitizenGrid allows different project teams to advertise, host and deploy worker nodes for their volunteer computing applications. The project team registers the VAS project within CitizenGrid and uploads a VAS contextualised micro-CernVM as the image to be used for the VAS worker node. CitizenGrid has been extended to include the option to personalise worker nodes by automatically embedding the \textit{team-id} into the contexualization process that takes place when starting a VAS compute node.

\section{Usage Scenarios} \label{UseCases}
Through the integration of the CitizenGrid, LiveQ and VAS platforms we have demonstrated how volunteer computing and volunteer thinking can be integrated within a single application. This integration benefits both the users of the VAS game and the community of volunteer computing providers who can join groups through CitizenGrid and target their computing power to help corresponding teams within VAS. However, this is just one example of how the integration of these tools can help volunteers contributing both thinking and computational power. In this section we look at three different usage scenarios facilitated through the integration of the tools described in this paper. In particular, we look at examples of how the integration of CitizenGrid and LiveQ can provide a generic platform for managing applications, groups of volunteers and the distribution of computational tasks to computing volunteers.

\subsection{Aggregation of donated resources}
\begin{figure}[h]
 \centering
\fcolorbox{black}{white}{ \includegraphics[width=80mm,height=40mm]{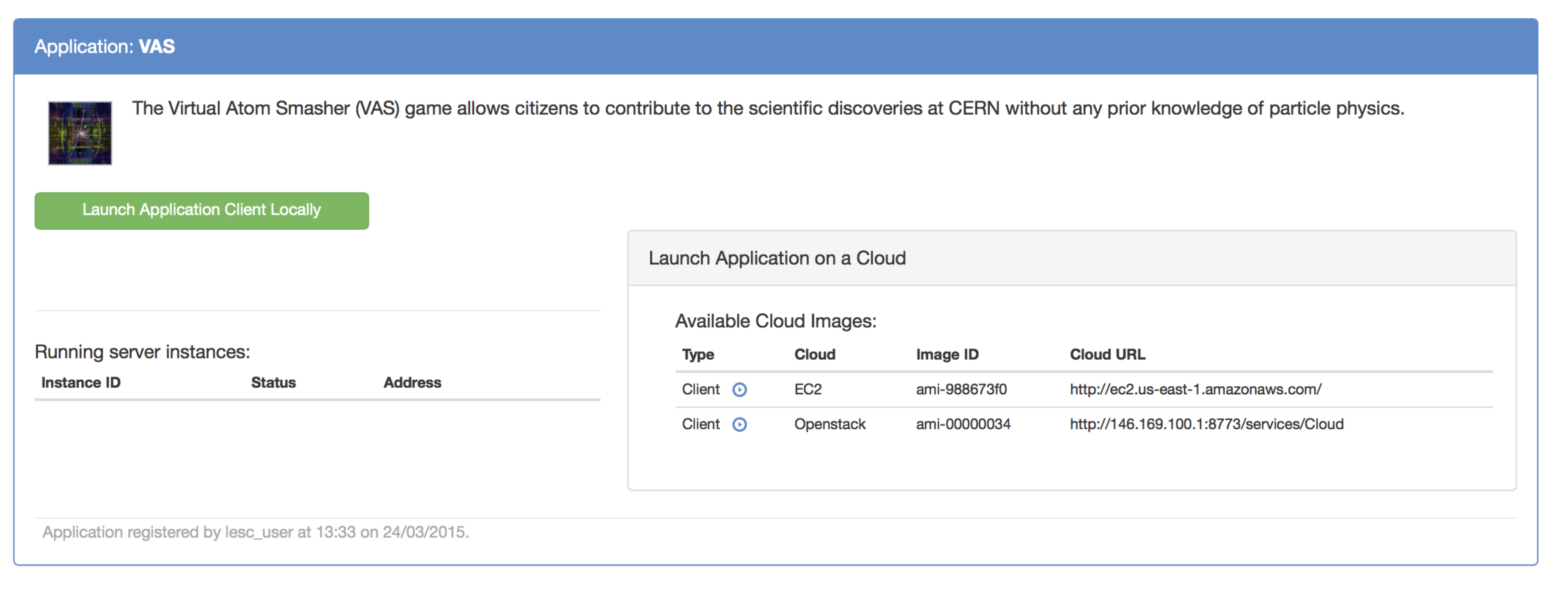}} 
\caption{CitizenGrid: VAS application client details.}
\label{fig:CGappdetails} 
\end{figure}

In Citizen Science, volunteers can choose different ways to contribute their knowledge or computing resources to a wide range of available scientific projects. In the case of volunteer computing, this requires that VC projects are set up in such a way that they can take advantage of using a number of distributed computational resources to service their computational requirements, for example, their local machine(s) or remote cloud computing resources. Existing VC projects generally rely on volunteers' personal computers and do not make use of cloud computing infrastructure. However, CitizenGrid offers volunteers the ability to use their local machine(s) or remote cloud computing resources to undertake VC tasks for a project of their choice. CitizenGrid allows the use of any distributed task management middleware such as BOINC or LiveQ for results aggregation at the application's server. Figure~\ref{fig:CGappdetails} shows the CitizenGrid interface displaying the application details for the VAS game. It shows two registered cloud images targeting either the OpenStack private cloud or Amazon EC2 public cloud platforms. The user can choose to start one or more instances on one of these platforms depending on what platforms they have access to.

Additionally, by integrating CitizenGrid and LiveQ we can provide a comprehensive framework for enabling users to offer their resources in a targeted manner and for application providers to efficiently aggregate a potentially large pool of geographically distributed resources to service their project's computational requirements. The group capabilities within CitizenGrid that allow "team-based" participation in projects offer further capabilities for the CCS projects to use team-based approaches and team-focused incentives within their applications. Volunteers may join different teams and contribute their resources to different projects collaboratively or individually giving them flexibility in how their resource capacity is used, via a single interface.

\subsection{Meeting the demand for real-time computation}
In games such as VAS, real-time computation is important to ensure that players receive feedback in a reasonable time when the underlying system is undertaking tasks that are computationally intensive. The VAS game interface could be replaced by a range of other computationally intensive scientific applications which require real-time computing responses, e.g., a disaster model mapping application that requires huge computing resources in a short span of time to generate a real time disaster map based on real time data to predict and understand the effects of a disaster. Two very popular Citizen Science human computation projects, Foldit~\cite{Khatib2012,Foldit} and Eyewire~\cite{Eyewire2015, Eyewire}, that currently attract up to four thousand volunteers everyday, are an ideal example of the challenges of managing the demand for real-time computation. The computational resources required in these games are significant and the projects currently use dedicated servers. In such an environment, handling a sudden increase in demand from users is a major challenge. Game operators face their resources not being able to stand up to the requirements for computation and ultimately crashing under an unmanageable load from the increased user requirements. On the other hand, when usage is low, if a dedicated resource pool is being used to operate the game, resource capacity is likely to be wasted. While cloud computing infrastructure is one possible way of handling changes in demand, this is likely to be costly over a long period of time and impractical for a freely available scientific game. By integrating with a volunteer computing environment, game operators can take advantage of contributed resources made available by VC providers. An environment like CitizenGrid offers a way to pool VC providers and build a community of volunteers who can respond when an application has a need for more computing power. This is something that can be challenging to handle in an open environment where it may be difficult to get access to volunteers' resources at the times when they are needed.

\subsection{Platform for promoting STEM} 
The final area where we see the integration of CitizenGrid, LiveQ and VAS as an ideal model to support other volunteer computing and volunteer thinking applications is in the promotion of science and technology education. The CitizenGrid-VAS platform can be used as a teaching tool in schools and universities for informal learning. In recent years, a number of do-it-yourself  open-source game frameworks, such as RedWire~\cite{Redwire2015}, have been designed and implemented and these could be used in place of the VAS game to offer alternative game environments to volunteers. Such platforms can be used to design scientific or educational games and can take advantage of the CitizenGrid and LiveQ integrated platform to offer a way to promote their games to potential users and access volunteer computing power to handle any computationally intensive aspects of the games. Using such tools, researchers/game developers can also write exciting standalone educational games which can subsequently be integrated with volunteer computing projects using the standard interfaces provided by the CitizenGrid and LiveQ platforms.

\section{Related Work}\label{Related}
The concept of combining volunteer gaming along with volunteer computing, where volunteers donate their computing resources to run simulations or other tasks, has been suggested in the literature~\cite{Cusack2006,Cusack2008}. However, these games solve only specific problems and are tightly coupled with volunteers' human computation tasks. In the last decade, the importance of games in human computation (volunteer thinking) has been confirmed by many studies~\cite{Yuen2009, Bowser2013a, Oliveira2017}. There have been a number of human computation games that have been implemented to solve some complex problems~\cite{Curtis2014}, for example, the Wildfire Wally game was designed to solve graph search problems~\cite{peck2007wildfire}. These games are casual games, and there are no scientific and technical learning components associated with them~\cite{peck2007wildfire, Cooper2010, Toth2007, Simko2011}. However, there is significant complexity involved in the design of these games to make the most of the potential of human computation~\cite{Cooper2010}. In addition to this, the complexities increase when these games are integrated with volunteer computing tasks. In current literature, only a small number of platforms exist that act as directories to host volunteer computing projects~\cite{WCG2015, Yadav2015CitizenGrid, Boinc}. A popular platform with users is the IBM World Community Grid~\cite{WCG2015} which currently hosts five active projects that use the BOINC~\cite{Boinc} middleware for distributed task management.  There are a number of client-server based volunteer computing middleware platforms that have been developed in last two decades, e.g., BOINC~\cite{Boinc}, Co-pilot~\cite{Copilot}, Cosm~\cite{Cosm1995}, and Bayanihan~\cite{Luis1999}.  A few successful projects, for example, Folding@home~\cite{Folding2000,Beberg2009} and Distributed.net~\cite{Distributednet} use the Cosm networking libraries for distributed networking. However, the majority of volunteer computing projects that have started in last few years use BOINC~\cite{Boinc}. BOINC is a distributed batch processing system. BOINC clients, which run on volunteers' computers, can pull tasks from the BOINC server when they are idle in order to process them locally. The volunteer's computer stores a batch of processed tasks before sending them back to the project server.  Co-pilot~\cite{Copilot} is another distributed task management system that is designed for CERN distributed computing projects, which acts as a \textit{gateway} between CERN's grid-infrastructure and volunteer resources. Co-pilot allows the distribution of CERN worker nodes onto both cloud infrastructure and volunteers' computers. However, none of the above-mentioned distributed task management systems is designed for real-time interaction, making them unsuitable for an interactive game environment where a player's next-move depends on the outcome of their previous move.
  To the best of our knowledge, interactive games where real-time VC task distribution and management are combined are not described in the CCS literature.  The model of integrating CitizenGrid with the LiveQ framework for handling distribution of volunteer computing tasks to volunteers' machines provides the flexibility of integrating a wide variety of scientific games to a distributed computing infrastructure which requires real-time computation to be carried out by volunteers' resources.

\section{Conclusions and Future Work}\label{Conclusion}
In this paper, we have presented a CCS deployment scenario which combines volunteer computing and thinking tasks within a single project.  We have also shown how three existing platforms CitizenGrid, LiveQ  and VAS can be linked to form an integrated CCS environment.  CitizenGrid is a middleware platform for deploying VC clients either using virtual machine images on volunteers' computers or on cloud infrastructures such as Amazon EC2. The Virtual Atom Smasher (VAS) is an online particle physics game and LiveQ is a real-time task distribution and management system.  

Citizengrid-VAS, the platform resulting from the integration of these three tools, demonstrates how a game-based collaborative volunteer computing platform can be integrated with a citizen science game. The platform allows easy deployment of a scientific and educational game that requires the use of volunteers' donated computing resources (free CPU cycles) in real-time. The CitizenGrid and LiveQ aspect of the integrated platform is flexible and could be used in a variety of other CCS project scenarios. As part of our future work we are planning to extend this collaborative platform to include the following new functionality: (1) A  resource request management interface on the CitizenGrid platform, which allows game players to initiate a request for volunteer resources. Volunteers then decide to choose to provide their computing resources to a particular game player. (2) Decoupling of LiveQ, CERN VM and the VAS game interface. This will provide more flexibility to a game designer to use LiveQ and VAS with other platforms.

\section{Acknowledgements}
The authors would like to thank the European Commission for funding the citizen cyberlab project (Grant number 317705), under their 7th Framework Programme, through which much of the work described here was undertaken. We would also like to thank members of the VAS, LiveQ and CitizenGrid teams for their support of this work.

\ifCLASSOPTIONcaptionsoff
  \newpage
\fi



\bibliographystyle{IEEEtran}
\bibliography{IEEEabrv,FINAL VERSION.bbl}

%

\begin{IEEEbiography}[{\includegraphics[width=1in,height=1.25in,clip,keepaspectratio]{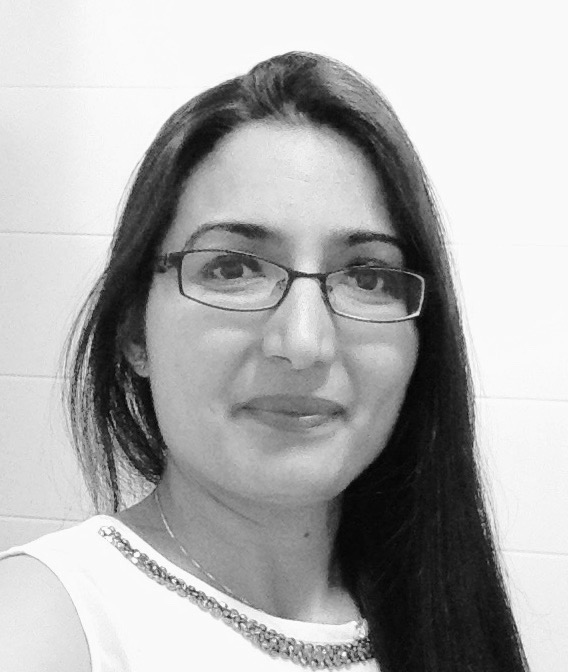}}]{Poonam Yadav}
is a research and teaching associate at the Computer Laboratory, University of Cambridge, UK.  She received the PhD degree in Computing from Imperial College London, in 2011 and M.Tech from IIIT, Allahabad, India, in 2007.  She is a recipient of UK-India Education and Research Initiative (UKIERI) PhD Award and has worked on various NERC, TSB, EU, EPSRC, IBM and Microsoft funded research projects and an author of over 30 papers in Distributed Systems, Social Computing, Sensor Systems and IoT. She is currently the Chair of ACM-W UK professional chapter and a member of ACM, IEEE and BCS.
\end{IEEEbiography}
 
\begin{IEEEbiography}[{\includegraphics[width=1in,height=1.25in,clip,keepaspectratio]{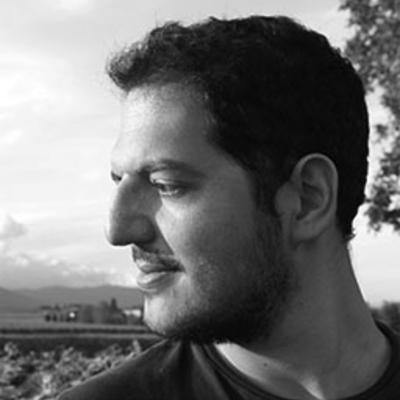}}]{Ioannis Charalampidis} is working as a Frontend Engineer since August 2016.  Previously he was OpenLab  fellow  at CERN, where he worked as an Architect and Software Engineer in PH/TH group. He is the lead developer of Virtual Atom Smasher game prototype and have published  a number of research papers on CernVM, cloud computing and virtualization.  
\end{IEEEbiography}

\begin{IEEEbiography}[{\includegraphics[width=1in,height=1.25in,clip,keepaspectratio]{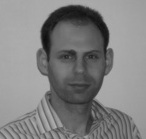}}]{Jeremy Cohen} is a Research Fellow in the Department of Computing at Imperial College London. He received the Ph.D. degree in computer science from Imperial College, University of London, London, UK and his research focuses on e-Science and the development of tools, services and applications to simplify access for domain scientists to a range of different computational infrastructure.
\end{IEEEbiography}
\begin{IEEEbiography}[{\includegraphics[width=1in,height=1.25in,clip,keepaspectratio]{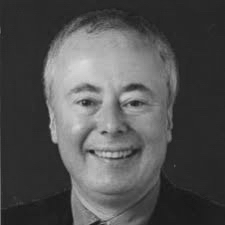}}]{John Darlington} is a professor  and  the head of the Social Computing Group in the Department of Computing  at Imperial College London. Professor Darlington has a long and distinguished track record both in the development of novel software technologies and in the creation of facilities to improve the accessibility and ease of use of computational resources. His work included pioneering developments in functional programming languages, program transformation, functional skeletons, co-ordination forms (later adopted by Google as map/reduce) and component-based application development frameworks. Facility developments have included the founding and operation of the Imperial College Fujitsu Parallel Research Centre, the Imperial College Parallel Computing Centre, the London e-Science Centre and the Imperial College Internet Centre. Professor Darlington's current interests include the use of Cloud computing to support a variety of public and civic Internet services and applications and the economic and social issues surrounding their successful use.
\end{IEEEbiography}
\begin{IEEEbiography}[{\includegraphics[width=1in,height=1.25in,clip,keepaspectratio]{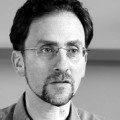}}]{Francois Grey}  is a physicist with a passion for citizen science. Since September 2014, he is Invited Professor at the University of Geneva and Coordinator of Citizen Cyberlab (CCL), a partnership between CERN, the United Nations Institute for Training and Research and the University of Geneva. He is co-author of 7 patent applications (5 granted) and over 100 scientific publications in peer-reviewed international journals.
\end{IEEEbiography}





\end{document}